\documentclass[letter,onecolumn]{jpsj3}%
\usepackage{amsfonts}
\usepackage{amsmath}
\usepackage{amssymb}
\usepackage{graphicx}%
\newcommand{\PR}[1]{Phys. Rev. B {\bf {#1}}}

\newcommand{\PRL}[1]{Phys.\ Rev.\ Lett. {\bf {#1}}}
\newcommand{\JPSJ}[1]{J.\ Phys.\ Soc.\ Jpn. {\bf #1}}

\newcommand{\slt}{$\raisebox{-0.6ex}{$\stackrel{<}{\sim}$}$}
\newcommand{\bvec}[1]{\mbox{\boldmath $#1$}}
\newcommand{\D}{\delta }
\newcommand{\vq}{\bvec{q}}

\newcommand{\vk}{\bvec{k}}

\newcommand{\Deta}{\Delta\!\eta}
\newcommand{\eq}[1]{eq.~(\ref{#1})}
\newcommand{\fig}[1]{Fig.~\ref{#1}}

\newcommand{\be}{\begin{equation}}
\newcommand{\ee}{\end{equation}}
\newcommand{\bea}{\begin{eqnarray}}

\newcommand{\eea}{\end{eqnarray}}
\newcommand{\bean}{\begin{eqnarray*}}

\newcommand{\bfi}{\begin{figure}}
\newcommand{\efi}{\end{figure}}
\newcommand{\bc}{\begin{center}}
\newcommand{\ec}{\end{center}}

\newcommand{\lsim}{ < \kern -11.8pt \lower 5pt \hbox{$\displaystyle \sim$}}
\newcommand{\gsim}{ > \kern -12pt   \lower 5pt   \hbox{$\displaystyle \sim$}}

\begin{document}

\title{Possible Quasi-One-Dimensional Fermi 
Surface in La$_{2-x}$Sr$_{x}$CuO$_4$}
\author{Hiroyuki Yamase$^{1}$ and Hiroshi Kohno$^{2}$}

\inst{$^{1}$Institute for Solid State Physics, University of Tokyo, 
 7-22-1 Roppongi, Minato-ku, Tokyo 106-8666 \\ 
$^{2}$Graduate School of Engineering Science, 
Osaka University, Toyonaka, Osaka 560-8531}

\recdate{September 14, 1999}

\abst{To reconcile the two experimental findings on 
La$_{2-x}$Sr$_{x}$CuO$_4$, 
namely, Fermi surface (FS) observed by angle-resolved photoemission 
spectroscopy and sharp incommensurate magnetic peaks 
by neutron scattering, 
we propose a picture 
that a quasi-one-dimensional FS (q-1dFS) is realized in each 
CuO$_{2}$ plane whose q-1d direction alternates along the 
$c$-axis.} 

\kword
{LSCO, $t$-$J$ model, incommensurate antiferromagnetism, 
Fermi surface, ARPES, neutron scattering}

\maketitle

Early studies by neutron scattering 
revealed the material dependence of 
low-energy spin excitations in 
high-$T_{\rm c}$ cuprates: 
La$_{2-x}$Sr$_{x}$CuO$_4$ (LSCO)\cite{yamada} in the metallic region    
shows  sharp incommensurate antiferromagnetic (IC-AF) peaks at wavevectors 
$(\pi \pm 2 \pi \eta, \pi)$ and $(\pi, \pi \pm 2 \pi \eta)$,   
where $\eta >0$ is called incommensurability, while in 
YBa$_{2}$Cu$_{3}$O$_{6+y}$ (YBCO)\cite{rossat-mignod,rossat-mignod2} 
there is a single broad commensurate 
structure at wavevector $(\pi, \pi)$.   
 It was pointed out theoretically\cite{si} that 
such material dependence could be understood 
in terms of the shape of the Fermi surface (FS): 
in the calculation based on the $t$-$J$ model\cite{tanamoto}, 
a diamond-shaped \lq electron-like' FS centered at $(0,0)$ 
suggested  an IC-AF fluctuation, 
while an almost circular \lq hole-like' FS centered at $(\pi, \pi)$ 
suggested a commensurate one. 
  The former has been presumed to be applicable 
to LSCO, and the latter to YBCO.

  However, angle-resolved photoemission spectroscopy (ARPES)\cite{ino} 
performed recently on LSCO has revealed some segments of the FS,  
which looks more consistent with the FS of YBCO-type rather than LSCO-type,  
at low doping rates, $\D$\cite{xD}. 
  This experimental finding 
has raised a serious question as to 
whether  low-energy spin excitation 
in LSCO can be understood in terms of fermiology.

  There is an interesting indication\cite{kuroki}, 
based on the fluctuation exchange (FLEX) approximation 
on the Hubbard model, that even with YBCO-type FS,  
the IC peak can be reproduced, 
suggesting that fermiology is sufficient 
for understanding the IC-AF fluctuation.  
However,  such IC peaks are not isolated but are connected 
with each other 
with substantial spectral weight at wavevector $(\pi, \pi)$, 
resulting in an essentially commensurate peak with an IC substructure.  
Thus, YBCO-type FS consistent with ARPES data seems 
incompatible with the sharp 
IC-AF peaks observed by inelastic neutron scattering.

  How can we reconcile these two experimental findings, namely those by 
ARPES and neutron scattering?
  One solution may be to abandon the connection of the IC-AF fluctuation 
to fermiology. 
  Actually, the currently growing 
picture\cite{tranquada1,tranquada2,tranquada3}  
of \lq charge stripe'  
may offer 
another route to the IC peaks without recourse to fermiology. 

  In this paper, however, we point out that we can 
construct a consistent picture between fermiology 
and sharp IC-AF peaks in LSCO  if we assume a  
quasi-one-dimensional FS (q-1dFS), not a two-dimensional FS (2dFS),  
in each CuO$_{2}$ plane.  
This assumption is based on our recent theoretical result\cite{yamase3} 
that the 2d $t$-$J$ model treated in the mean-field approximation 
has an instability towards forming a q-1d band. 
(Remarkably, this instability occurs even in homogeneous charge 
distribution, and  we assume this charge homogeneity in this paper.)  
 Deferring the details of the microscopic origin of the 
q-1d band elsewhere\cite{yamase2},   
we here show 
that the FS observed by ARPES\cite{ino} 
can be regarded as consisting of 
two orthogonal q-1dFSs; we can fit the former by the latter 
at each doping rate. 
  We then calculate the doping dependence of $\eta$, 
which is found to improve quantitatively the result calculated in the 
previous scheme\cite{tanamoto}. 
 We also explore the effects of orthorhombicity of the crystal structure 
on the wavevector of IC-AF, motivated by recent  elastic neutron scattering 
studies\cite{kimura,lee}.

Let us first assume 
that  either of two kinds of q-1dFSs, 
q-1dFS(x) or q-1dFS(y), 
is realized in each CuO$_{2}$ plane and is stacked alternately along 
the $c$-axis as shown in \fig{q1dfs along c}. 
This assumption of alternate stacking     
may be justified by the fact 
that the q-1d band couples to 
the soft phonon mode\cite{thurston} related to the instability toward the  
low-temperature tetragonal (LTT) structure.

We first show that our q-1dFS picture does not conflict with ARPES 
data\cite{ino}  that imply a 2dFS. 
Since there is a small, but not negligible, interlayer hopping integral, 
$\tilde{t}_{z}$\cite{pikett}, 
two dispersions, $\xi_{\vk}^{\rm A}$ and $\xi_{\vk}^{\rm B}$, 
each forming q-1dFS(x) and 
q-1dFS(y), respectively, hybridize to become 
\be
\xi_{\vk}^{\pm}=\frac{\xi_{\vk}^{\rm A} + \xi_{\vk}^{\rm B}\pm 
\sqrt{(\xi_{\vk}^{\rm A} - \xi_{\vk}^{\rm B})^2 +4 \epsilon_{\vk}^2}} {2}, 
\label{hybridization}
\ee
where 
\be
\epsilon_{\vk}=\tilde{t}_{z} \cos \frac{k_{x}}{2} \cos \frac{k_{y}}{2} 
	\cos \frac{k_{z}}{2}.   
\ee
The form factor, $\cos \frac{k_{x}}{2} \cos \frac{k_{y}}{2} 
\cos \frac{k_{z}}{2}$, 
comes from 
the fact that Cu sites on adjacent 
CuO$_{2}$ planes are  relatively displaced by 
$[1/2,\,1/2,\,1/2]$. 
The resulting FS consists of the 
\lq outer FS' formed by  $\xi_{\vk}^{-}$  
and the \lq inner FS' by $\xi_{\vk}^{+}$ 
as shown in \fig{superposition}, where 
the band width of 
 $\epsilon_{\vk}$ is taken to be about 0.1 times that of the
 $\xi_{\vk}^{\rm A}$ (or $\xi_{\vk}^{\rm B}$). 
Note that $\epsilon_{\vk}=0$ at $(\pi,\,0)$ and $(0,\,\pi)$, where 
no dispersion appears  along $k_{z}$. 
Since the ARPES spectrum has a very broad peak   
 even near $(\pi,\,0)$ and $(0,\,\pi)$, the effect of the dispersion 
along $k_{z}$  
will be sufficient to blur the \lq inner FS' and the 
\lq outer FS' except near  $(\pi,\,0)$ and $(0,\,\pi)$, if the ARPES 
spectrum is somewhat integrated along the  $k_{z}$-direction. 
 This interpretation is consistent with the ARPES data\cite{ino}, in which 
the FS is always detected  near $(\pi,\,0)$ and $(0,\,\pi)$ 
for $0.05 \leq \D \leq 0.30$, 
but not near $(\pi/2,\,\pi/2)$ especially below $\D \sim 0.15$. 
Therefore, we propose that the FS observed by ARPES 
comes from the q-1dFS(x) and the q-1dFS(y). 
We note here a recent finding in ARPES\cite{shen} that momentum   
distribution, $n(\vk\, ; \,\Lambda)$, depends on the range,
 $\Lambda$, of energy integration near the Fermi energy. 
This dependence will also be understood as coming from effects of 
$\tilde{t}_{z}$: 
for $\Lambda < \left | \tilde{t}_{z}\right |$, $n(\vk\, ; \,\Lambda)$ 
reflects effects of 
the dispersion along $k_{z}$, while it does not for 
$\Lambda \gg \left | \tilde{t}_{z}\right | $.   
(These discussions hold even if we consider $d$-wave singlet 
pairing in \eq{hybridization}.)

In order to determine the q-1d dispersions, 
$\xi_{\vk}^{\rm A}$ and $\xi_{\vk}^{\rm B}$, we 
use the previous mean field theory\cite{tanamoto} (MFT)  
of the 2d $t$-$J$ model on a square lattice.  
We set temperature as $T=0.01 J$ 
where the thermal smearing of the FS can be neglected. 
We take 
the parameters as $t/J=4$ and $t\,^{'}/t=-1/6$, 
where $J>0$ is the nearest neighbor (n.n.) superexchange coupling, 
$t$ is the first n.n. hopping integral and $t'$ is the second one. 
These parameters are chosen so as to reproduce  the observed FS 
at $\D =0.30$ where the whole shape of the FS has been clarified\cite{ino}.   
Previously\cite{tanamoto},  the dispersion for  (spin) 
fermions was given by $E_{\vk}=\sqrt{\xi_{\vk}^2+ |\Delta_{\vk}|^2}$, 
where $\Delta_{\vk}=- \frac{3}{4}\Delta_{0}(\cos k_{x} - \cos k_{y} )$ 
($\Delta_{0}$ is the singlet order parameter) and 
\be
\xi_{\vk}= - 2\left(F_{x} \cos k_{x} + F_{y} \cos k_{y} 
+ F\,^{'}_{\parallel} \cos (k_{x}+k_{y}) 
    +  F\,^{'}_{\perp} \cos (k_{x}-k_{y}) \right) -\mu , 
\ee
with $F_{x}=F_{y}$, $F\,^{'}_{\parallel}=F\,^{'}_{\perp}$ and  
$\mu$ being the  chemical potential.  
The values of $F_{x},F\,^{'}_{\parallel}$, $\mu$ and $\Delta_{0}$ 
were determined numerically 
by minimizing the free energy. 
The resulting FS was used for  discussing LSCO\cite{tanamoto}. 
We call this FS \lq previous FS', because  
recent ARPES\cite{ino} has revealed   
that the \lq previous FS' is incompatible with  the 
observed FS especially for low $\D$.  
Thus, in this paper, we reduce the value of $F_{y}$ as 
\be
F_{y}=\alpha F_{x} , 
\ee
for the q-1dFS(x), retaining for 
the other parameters, $F_{x},F\,^{'}_{\parallel}, F\,^{'}_{\perp}, \mu$ 
and $\Delta_{0}$, the values of the  previous MFT\cite{tanamoto}. 
(For the q-1dFS(y) we take $F_{x}= \alpha F_{y}$.)  
We choose the value of $\alpha$ to adjust 
the q-1dFS(x)  near $(0,\,\pi)$ to the observed FS segments.  
The obtained values are $\alpha = 0.75,\,0.78,\,0.85,\,0.92$ and $1.0$ 
for $\D=0.05,\,0.10,\,0.15,\,0.22$ and $0.30$, respectively: 
the band anisotropy decreases with doping and the 2dFS is realized 
at $\D=0.30$.

Next, we study the wavevector of IC-AF fluctuation in the state with the
q-1dFS. 
For simplicity,  
we neglect the interlayer 
coupling, $\tilde{t}_{z}$, and treat the q-1dFS(x) and the q-1dFS(y) 
independently. 
Since  the energy scale of the IC-AF fluctuation is about $0.02J$ 
($2 \sim 3$ meV) experimentally\cite{yamada},    
which is much smaller than $J$, 
it will  be reasonable  to estimate the wavevector of IC-AF fluctuation from 
the peak position of the static  magnetic susceptibility, 
$\chi_{0}(\vq)$, for non-interacting fermions. 
 In \fig{chi0}, we plot $\chi_{0}(\vq)$ for the q-1dFS(x) along 
$(\pi,\,\pi/2)\rightarrow (\pi,\,\pi)\rightarrow 
(\pi/2,\,\pi)$ at $\D=0.15$ as a typical example.  
We see two IC peaks whose heights are comparable: 
the IC peak along $(\pi,\,\pi/2) \rightarrow (\pi,\,\pi)$ is  
located at  $(\pi,\, \pi - 2\pi\eta^{y})$ and 
further deviates from $(\pi,\,\pi)$ relative to that for the 
\lq previous FS', 
whereas  the other peak at $(\pi -2\pi\eta^{x},\,\pi)$ 
remains at almost the same position as 
that for the \lq previous FS'.  
(Note that IC peaks are 2d-like rather than 1d-like 
even in the state of the q-1dFS.)
For the q-1dFS(y), $\chi_{0}(\vq)$ 
is obtained by the 90$^{\circ}$ rotation from 
that for  the q-1dFS(x). 
Figure \ref{etadelta}(a) schematically shows 
the wavevector  at which $\chi_{0}(\vq)$ takes a peak for each q-1dFS.   
In \fig{etadelta}(b), $\eta^{x}$ and $\eta^{y}$ are shown 
as a function of $\D$: 
compared with the results for the \lq previous FS' (open triangles), 
the values of $\eta^{y}$ (filled  circles) become much larger especially 
in $0.10 \slt \D \slt 0.15$, while the values of $\eta^{x}$ (filled squares) 
are almost unchanged. 

Since inelastic neutron scattering data have not been analyzed 
in terms of two incommensurabilities,  
$\eta^{x}$ and $\eta^{y}$, comparison of the present results 
directly with the experimental $\eta$-$\D$ curve\cite{yamada} 
(indicated by crosses in Fig. 4(b)) cannot be done in a straightforward manner. 
(We leave as a future problem the  recently observed \lq diagonal IC peak'  
(located at $(\pi \pm 2\pi\eta,\,\pi \pm 2\pi\eta)$) 
at $\D=0.05$\cite{wakimoto}.)  
Nonetheless, the results for the q-1dFS seem valid in the following 
aspects. 
First, through the calculation of the ${\rm Im}\chi_{0}(\vq,\omega)$ 
at $\omega=0.01 {\rm J}$ 
and  $T=0.01 {\rm J}$, we find that in contrast to the 2dFS\cite{kuroki}, 
the q-1dFS yields sharp IC peaks even at low $\D$, which is 
consistent with experiments\cite{yamada}. 
A preliminary result for $\D=0.10$ is  shown in \fig{Imxo}. 
Second, the values of the $\eta^{y}$ for the 
q-1dFS(y) are closer to the experimental values than those for the 
\lq previous FS'.

There is another interesting 
phenomenon,  which we call the \lq shift', that 
the IC peaks do not lie exactly  
on the symmetry axes ($k_{x}=\pm \pi $ or 
$k_{y}=\pm \pi$) but deviate slightly from them as schematically 
shown in \fig{yshift}(a). 
 This \lq shift' is quantified by $\theta_{\rm Y}$ (or $\Deta$) as defined 
in the figure.   
 It has been reported that 
$\theta_{\rm Y} \sim 3.2^{\circ}$, $2.5^{\circ}$ (or 
$\Deta \sim 0.0067,0.0052$) for 
$\D=0.12$, $0.13$\cite{kimura}, respectively. (Hereafter, we use 
$\Deta$ as our notation.)  
 In the present context, the \lq shift' 
can be understood in terms of the anisotropy of the second n.n. transfer 
integrals in the orthorhombic crystal structure as follows.    

In the low-temperature orthorhombic  (LTO1) structure, 
CuO$_{6}$ octahedra tilt 
around the [110] axis (tetragonal notation) about 
$\theta= 3 \sim 4^{\circ}$\cite{radaelli} at $\D=0.10 \sim 0.15$. 
As a result, (i) the distance between the second n.n. Cu sites becomes longer 
along $[1\overline{1}0]$ while it gets shorter along $[110]$\cite{radaelli}, 
and (ii) the alternate buckling of O atoms occurs  
along $[1\overline{1}0]$ 
while the uniform buckling occurs along $[110]$. Hence, the 
wave function of the Zhang-Rice singlet\cite{zhang} becomes 
spatially  anisotropic, which leads to an anisotropy 
between  $F_{\|}\,^{'}$ (parallel to $[110]$)  and  
$F_{\bot}\,^{'}$ (perpendicular to [110]), 
namely $|F_{\|}\,^{'}| > |F_{\bot}\,^{'}|$.  
The effect (i) will produce the anisotropy of about  $1 \sim 2 \%$,   
if we assume $F\,^{'} \propto t_{pp}$ 
with\cite{harrison} $t_{pp} \propto r^{-2}$ 
where $t_{pp}$ is the hopping 
integral between the n.n. O sites in a CuO$_{2}$ plane with distance $r$ 
(the  distance between the second n.n. Cu sites is $2 r$).  
As for the effect (ii), the degree of the anisotropy is 
expected to be at least more than 
$1\sim 2 \%$, if we assume 
$F_{\bot}\,^{'}/F_{\|}\,^{'} > (1-3.78 \tan ^2 \theta)$  
where the form on  the right-hand side 
was used to estimate  the effect of the (uniform) 
buckling of O atoms on the hopping integral\cite{bruce}. 
We thus put  
\be
F^{\,'}_{\perp} = \gamma F^{\,'}_{\parallel}\; (\gamma\leq 1). 
\ee

We first calculate $\chi_{0}(\vq)$ at $\D=0.15$   
for various $\gamma$.  
When $\gamma=1$, $\chi_{0}(\vq)$ takes maxima 
at $(\pi,\,\pi\pm 2\pi\eta^{y})$ and 
$(\pi\pm 2\pi\eta^{x},\,\pi)$ for the q-1dFS(x). 
($\eta^{x}=0.092$ and $\eta^{y}=0.113$.)    
With decreasing $\gamma$, the 
positions shift to 
$(\pi \mp  2\pi \Delta\!\eta^{x},\,\pi \pm 2 \pi\eta^{y})$ and 
$(\pi\pm 2\pi\eta^{x},\,\pi \mp 2\pi \Delta\!\eta^{y})$ 
where both $\Deta^{x}$ and $\Deta^{y}$    
are positive\cite{shift}.  
Such a shift is 
consistent  with 
\fig{yshift}(a).  
The \lq shift' for the q-1dFS(y) is obtained by reflecting that 
for the q-1dFS(x) about the $q_{y}=q_{x}$ axis, and  
is also consistent with \fig{yshift}(a).     
This is because, while the shape of q-1dFS(y)  
is rotated by $90^{\circ}$, 
the tilting axis is not. In \fig{yshift}(b), we 
show $\Deta^{x}$ and $\Deta^{y}$ as a function of $\gamma$ 
where both $\Deta^{x}$ and $\Deta^{y}$ increase with decreasing $\gamma$. 
At $\D=0.10$, we find that  
the value  of  $\Delta\!\eta^{x}$ is almost the same as 
that for $\D=0.15$  at each $\gamma$, while the value of 
$\Delta\!\eta^{y}$ becomes  about 1.5 times larger. 
Thus, at $\D \approx 0.12 \sim 0.13$, 
the values of $\Deta^{x}$ and $\Deta^{y}$ will be 
comparable  to the experimental values when the anisotropy  
of $F^{\,'}$ is $\sim 10 \%$ $(\gamma \sim 0.9)$. 
This value of anisotropy is reasonable 
because, as estimated in the previous paragraph, 
the anisotropy due to the effects (i) and (ii) is expected to be  
at least more than $2\sim 4$\% in total. 
Therefore, fermiology can also explain 
the \lq shift' semiquantitatively in terms of 
the anisotropy of the 
second n.n. transfer integrals $F\,^{'}$ caused by the LTO1 structure. 
In the LTT structure, 
the \lq shift' is not expected in the present context 
because there is no anisotropy in $F\,^{'}$.  
This prediction can be tested experimentally on Nd-doped LSCO.

We have proposed a q-1dFS scenario for LSCO.   
Whether the same scenario applies to other high-$T_{c}$ cuprates,  
such as YBCO and Bi2212, is 
one of the questions we are trying to answer\cite{yamase2}. 
In addition, any relation of the present q-1dFS to the \lq charge stripe'  
picture\cite{tranquada1,tranquada2,tranquada3} has not been obtained 
in our current analysis and is  left as a future 
problem.  
As a test for the present picture, 
observations of (i) both \lq outer FS' and \lq inner FS', or 
(ii) both $\eta^{x}$ and $\eta^{y}$ will be crucial.   
As for (ii), however, we note some possibilities such as 
(a) $\eta^{x}$ and $\eta^{y}$ are too close compared to the peak width, 
and (b) IC-AF with either $\eta^{x}$ or $\eta^{y}$ develops due to the 
instability toward static IC-AF ordering, 
or due to the coupling\cite{miyake,yamase} 
to a possible vertical (or horizontal) 
charge stripe.

In summary, we have proposed, for LSCO, a scenario 
that a q-1dFS is realized in each CuO$_{2}$ plane whose 
q-1d direction alternates along the $c$-axis. We have shown that  
the observed FS can be understood as two orthogonal q-1dFSs. 
On the basis of fermiology, we have also shown that   
(i) IC-AF fluctuation has two 
incommensurabilities, $\eta^{x}$ and $\eta^{y}$, 
and (ii) the \lq shift' of sharp IC peaks 
from the symmetry axis can be understood as 
coming from the anisotropy of the second n.n. hopping integral 
in the LTO1 structure.  
As a test of the present picture, observations 
of both \lq outer FS' and \lq inner FS' will be conclusive.

We thank Dr. A. Ino, Dr. K. Kuroki and especially Professor H. Fukuyama  
for fruitful discussions. 
H. Y. also thanks Dr. H. Kimura and Mr. T. Yoshida 
for stimulating discussions.    
This work is supported by a Grant-in-Aid for Scientific Research from 
Monbusho.

\newpage

\bfi[t]
\bc
\includegraphics[width=10cm]{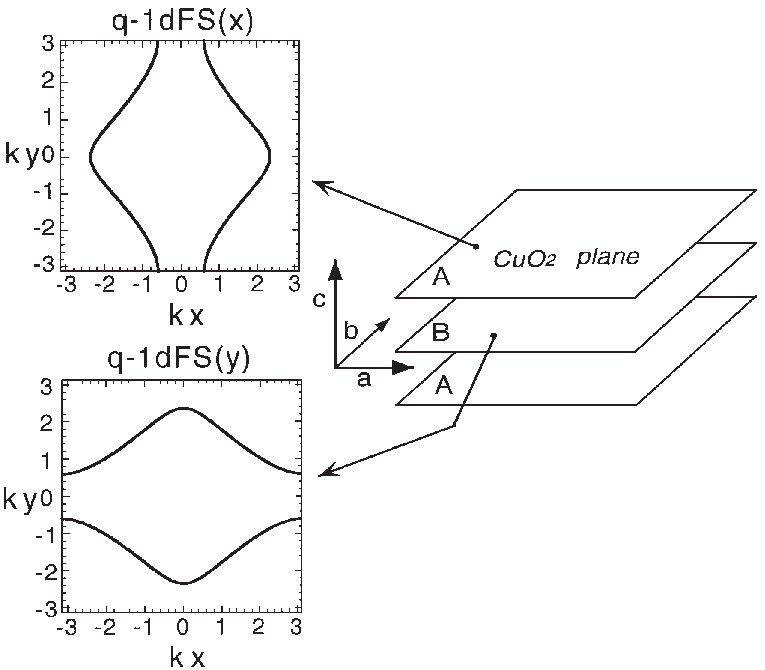}
   \caption{Proposed quasi-one-dimensional (q-1d) FSs, 
q-1dFS(x) and q-1dFS(y), and 
their alternate stacking along the c-axis. } 
    \label{q1dfs along c}
\ec
\efi

 \bfi
 \bc
\includegraphics[width=8cm]{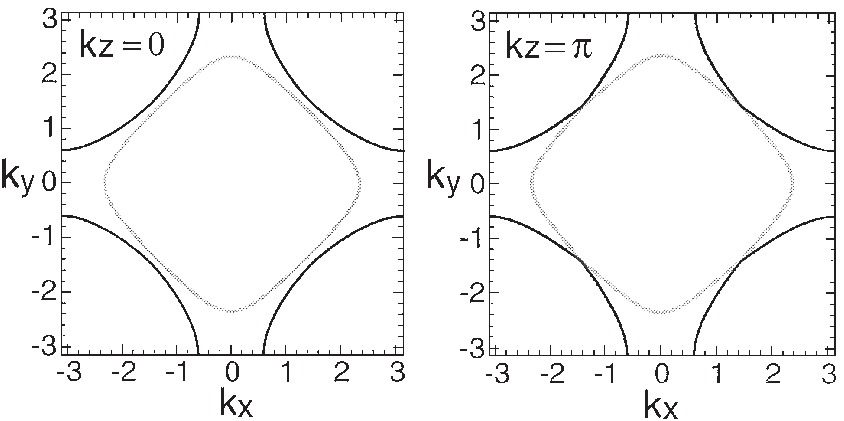}
   \caption{The theoretical FS at $k_{z}=0$ and $k_{z}=\pi$. The FS  
consists of the \lq outer FS', shown by the solid line, and 
the \lq inner FS', shown by the gray line.  
}
    \label{superposition}
    \ec
\efi

\bfi
 \bc
\includegraphics[width=8cm]{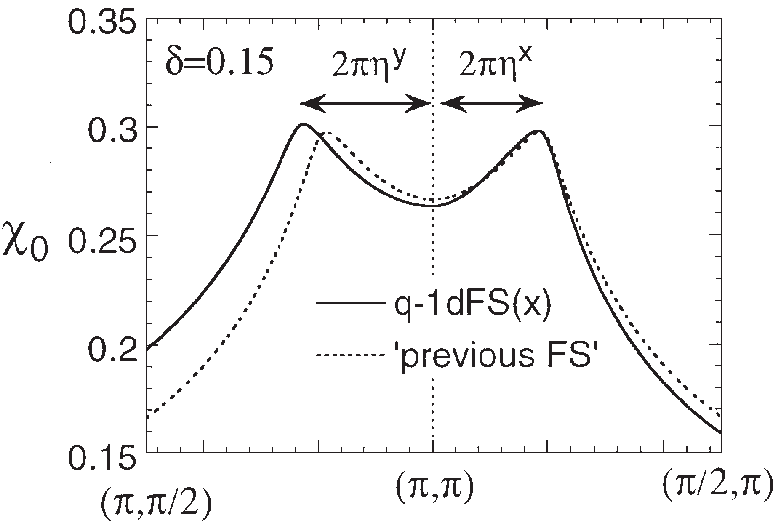}
   \caption{The $\chi_{0}(\vq)$ for the q-1dFS(x) 
at  $\D=0.15$. 
The peak positions of $\chi_{0}(\vq)$ are measured 
from $(\pi,\,\pi)$ as $\eta^{x}$ or $\eta^{y}$.  
For comparison, the result for the  \lq previous FS' (see the text) is  
shown by a dashed line.} 
    \label{chi0}
    \ec
\efi

\bfi
 \bc
\includegraphics[width=10cm]{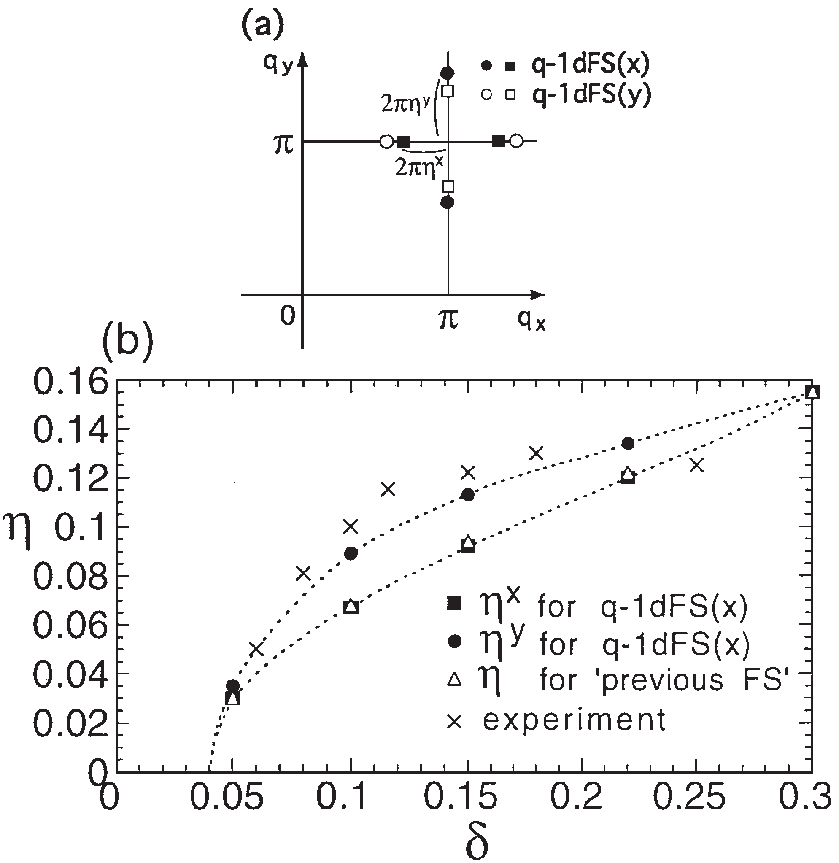}
   \caption{(a) Schematic figure of IC peak positions for 
the q-1dFS(x) (filled symbols) and the q-1dFS(y) (open symbols). 
Note that the positions  
for the q-1dFS(y) are obtained by the $90^{\circ}$ rotation   
from those for q-1dFS(x).  
(b) The $\eta^{x}$ and $\eta^{y}$ for q-1dFS(x) as a function of $\D$. 
Dashed lines are guides for the eye. For comparison, 
the result for the \lq previous FS' and experimental values are also shown. 
}
    \label{etadelta}
    \ec
\efi

\bfi
 \bc
\includegraphics[width=8cm]{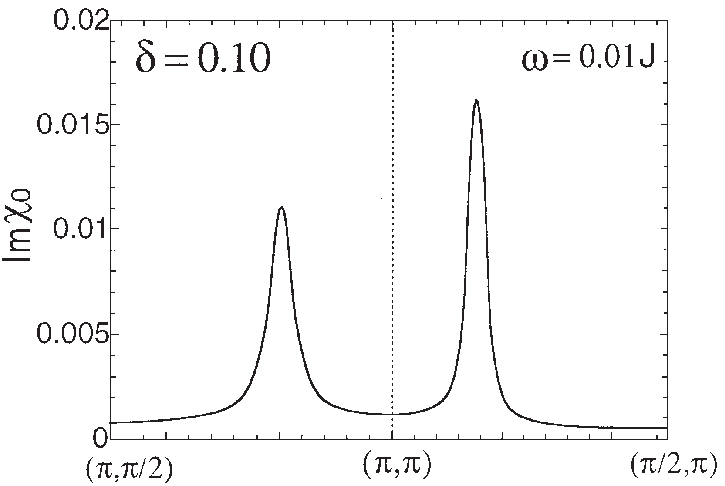}
   \caption{The sharp IC peaks of Im$\chi_{0}(\vq,\omega)$ at
   $\D=0.10$ and  $\omega=0.01 J$.  \hspace{7cm}}
    \label{Imxo}
    \ec
\efi

\bfi 
 \bc
\includegraphics[width=10cm]{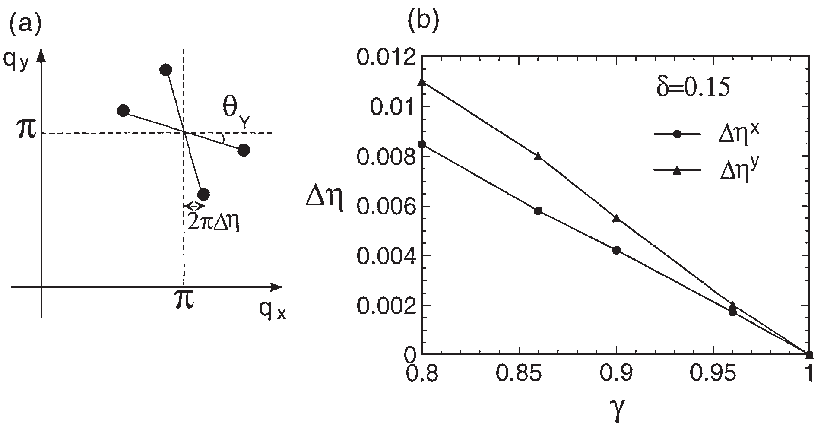}
 \caption{(a) Location of the elastic peaks observed by neutron scattering. 
They slightly shift away from the symmetry axis $(k_{x}=\pm \pi\; 
{\rm or}\; k_{y}= \pm \pi)$, and this \lq shift' is quantified 
by $\Deta$ or $\theta_{\rm Y}$.   
(b) The $\Deta^{x}$ and $\Deta^{y}$  
at $\D=0.15$ as a function of $\gamma$. For the definition of 
$\Deta^{x}$ and $\Deta^{y}$, see the text. }
 \label{yshift}
 \ec
\efi


\begin{thebibliography}{99}     

\bibitem{yamada} K. Yamada,  
C. H. Lee, K. Kurahashi, J. Wada, S. Wakimoto, 
                 S. Ueki, H. Kimura, Y. Endoh, S. Hosoya, G. Shirane, 
                 R. J. Birgeneau, M. Greven, M. A. Kastner and Y. J. Kim:
                 \PR{57} (1998) 6165 and references therein. 
\bibitem{rossat-mignod} J. Rossat-Mignod,  
J. P. Regnault, C. Vettier, 
P. Burlet, J. Y. Henry and G. Lapertot: Physica B {\bf 169} (1991) 58.
\bibitem{rossat-mignod2} J. Rossat-Mignod,  
J. P. Regnault, C. Vettier, 
P. Bourges, P. Burlet, J. Bossy, J. Y. Henry and G. Lapertot: 
Physica C {\bf 185 - 189} (1991) 86. 
\bibitem{si} Qimiao Si,  
Yuyao Zha, K. Levin and J. P. Lu: 
	\PR{47} (1993) 9055.  
\bibitem{tanamoto} T. Tanamoto,  
H. Kohno and H. Fukuyama: 
                   \JPSJ{62} (1993) 717. 
\bibitem{ino} A. Ino,  
C. Kim, T. Mizokawa, Z.-X. Shen. A. Fujimori, 
	M. Takaba, K. Tamasaku, H. Eisaki and S. Uchida: \JPSJ{68} (1999) 
	1496.  
\bibitem{xD} We consider $\D$ is equal to Sr$^{+2}$-content, $x$.  
\bibitem{kuroki}  K. Kuroki,  
R. Arita and H. Aoki: \PR{60} (1999) 9850.
\bibitem{tranquada1} J. M. Tranquada,  
B. J. Sternlieb, J. D. Axe, Y. Nakamura 
                    and S. Uchida: Nature {\bf 375} (1995) 561.    
\bibitem{tranquada2} J. M. Tranquada,  
J. D. Axe, N. Ichikawa, 
                    Y. Nakamura, S. Uchida and B. Nachumi: \PR{54}  
                     (1996) 7489.
\bibitem{tranquada3} J. M. Tranquada,  
J. D. Axe, N. Ichikawa, 
                     A. R. Moodenbaugh, Y. Nakamura and S. Uchida: \PRL{78}  
                     (1997) 338.

\bibitem{yamase3} H. Yamase, H. Kohno and H. Fukuyama: submitted to Physica C. 
\bibitem{yamase2} H. Yamase and H. Kohno: submitted to \JPSJ{}
\bibitem{kimura} H. Kimura: private communications. 
\bibitem{lee} Y. S. Lee,  
R. J. Birgeneau, M. A. Kastner, Y. Endoh, 
		S. Wakimoto, 
		K. Yamada, R. W. Erwin, S.-H. Lee and G. Shirane:
 \PR{60} (1999) 3643.
\bibitem{thurston} T. R. Thurston,  
R. J. Birgeneau, D. R. Gabbe,
 H. P. Jenssen, M. A. Kastner, P. J. Picone, N. W. Preyer, J. D. Axe,
 P. B\"{o}ni, G. Shirane, M. Sato, K. Fukuda and S. Shamoto: \PR{39}
 (1989) 4327. 
\bibitem{pikett} W. E. Pikett: Rev. Mod. Phys. {\bf 61} (1989) 433. 
\bibitem{shen} X. J. Zhou, P. Bogdanov, S. A. Kellar, T. Noda,
 H. Eisaki, S. Uchida, Z. Hussain and Z.-X. Shen: Science {\bf 286}
 (1999) 268. 
\bibitem{wakimoto} S. Wakimoto,  
G. Shirane, Y. Endoh, K. Hirota,
 S. Ueki, K. Yamada, R. J. Birgeneau, M. A. Kastner, Y. S. Lee,
 P. M. Gehring and S. H. Lee: \PR{60} (1999) R769. 


\bibitem{radaelli} P. G. Radaelli,  
D. G. Hinks, A. W. Mitchell,
 B. A. Hunter, J. L. Wagner, B. Dabrowski, K. G. Vandervoort,
 H. K. Viswanathan and J. D. Jorgensen: \PR{49} (1994) 4163. 
\bibitem{zhang} F. C. Zhang and T. M. Rice:  \PR{37} (1988) 3759. 
\bibitem{harrison} W. A. Harisson: {\it Electronic Structure and 
            the Properties of Solids} (Freeman, New York, 1980). 
\bibitem{bruce} B. Normand,  
H. Kohno and H. Fukuyama: \PR{53} (1996) 856. 
\bibitem{shift}  Strictly speaking, the values of $\eta^{x}$ and $\eta^{y}$ 
become larger by about $1 \sim 2$\% with decreasing $\gamma$. 


\bibitem{miyake} K. Miyake and O. Narikiyo: \JPSJ{63} (1994) 2042.
\bibitem{yamase} H. Yamase, 
H. Kohno, H. Fukuyama and M. Ogata: 
                 \JPSJ{68} (1999) 1082. 
\end{thebibliography}
\end{document}